\documentclass[aip,preprint,a4]{revtex4-1}
\usepackage{amsmath}
\usepackage{bm}
\usepackage{color}
\usepackage{epsfig}
\usepackage{float}
\usepackage{hyperref}

\begin{document}
\title{A theoretical model for linear and nonlinear spectroscopy 
of plexcitons}

\author{Chenghong Huang}
\author{Shuming Bai}\email{baishuming@iccas.ac.cn}
\author{Qiang Shi}\email{qshi@iccas.ac.cn}
\affiliation{Beijing National Laboratory for
Molecular Sciences, State Key Laboratory for Structural Chemistry of
Unstable and Stable Species, Institute of Chemistry, Chinese Academy
of Sciences, Zhongguancun, Beijing 100190, China}
\affiliation{University of Chinese Academy of Sciences,
Beijing 100049, China}

\begin{abstract}

We present a theoretical model to investigate the dynamics
and spectroscopic properties of a plexciton system
consisting of a molecular exciton coupled to a single 
short-lived plasmonic
mode. The exciton is described as a two-level system (TLS),
while the plasmonic mode is treated as a dissipative
harmonic oscillator. The
hierarchical equations of motion method is employed
to simulate energy transfer dynamics, absorption spectra,
and two-dimensional electronic spectra (2DES) of the system
across a range of coupling strengths. 
It is shown that increasing the exciton-plasmon
coupling strength drives a transition in the absorption
spectra from an asymmetric Fano line shape to a Rabi
splitting pattern, while coupling the TLS to intramolecular 
vibrational modes reduces the central dip of the absorption 
spectra and makes the line shape more symmetric.
The simulated 2DES exhibit distinct features
compared to those of a coupled molecular dimer, highlighting
the unique nonlinear response of plexciton systems. In
addition, a ``breathing mode” pattern observed in the strong
coupling regime can serve as a direct evidence of Rabi
oscillation.

\end{abstract}
\maketitle

\section{Introduction}
Plasmons are collective oscillations of electrons in metals
or metallic nanoparticles.\cite{shahbazyan13,tame13} They
produce highly localized electromagnetic fields that enable
strong interactions with nearby excitons, which are bound
states of electrons and holes in molecules or
semiconductors. When plasmons and excitons are brought close
together, they can couple to form hybrid states called
plexcitons. Plexcitons combine the tight field confinement
of plasmons with the discrete energy levels of excitons,
resulting in unique optical properties suitable for
applications ranging from solar cells to quantum information
technologies.\cite{torma15,li18,yu19,xiong21,wang22,kim23}

The exciton-plasmon coupled systems can be classified by
their individual damping rate and how fast energy exchanges
between them.\cite{luo19,kim23} In the weak coupling regime,
the plasmon damping rate is significantly higher than that
of the energy exchange between the exciton and plasmon. The
behavior of exciton is thus minimally influenced by plasmon
in such cases, while the emission rate of the quantum
emitter can be enhanced due to the Purcell effect,\cite{purcell95}
which have been observed in various 
systems.\cite{zhang15a,pelton15,badshah20,su21}

In the strong coupling regime, where the energy exchange
rate between the exciton and plasmon exceeds their
respective damping rates, the hybrid plexciton states 
are formed, resulting in the characteristic Rabi
splitting.\cite{schlather13,zhao22} This regime can be
achieved by carefully selecting the metal materials,
precisely controlling the shape and size of the metal
nanoparticles, and optimizing the coupling mechanisms
between the emitters and plasmons.

In recent decades, many experiments have demonstrated
peak splitting in linear spectroscopy for quantum emitters
strongly coupled to plasmonic nanoparticles. For instance,
\citeauthor{zengin13} studied TDBC J-aggregates both as thin
layers surrounding a silver nanorod and as sheets attached to
a silver nanoprism, observing peak splittings of $976~
\mathrm{cm^{-1}}$ and $2880~\mathrm{cm^{-1}}$ in dark-field
scattering spectra.\cite{zengin13,zengin15}. 
\citeauthor{chikkaraddy16} reported strong coupling of small
numbers of dye molecules to plasmonic nanocavities made of
gold nanoparticles on gold mirrors 
at room temperature.\cite{chikkaraddy16}
\citeauthor{santhosh16} observed vacuum Rabi splitting in
a single quantum dot coupled to plasmonic cavities formed by
a silver nanoparticle dimer, with peak splitting comparable
to that of multi-emitter plasmonic systems.\cite{santhosh16}.
Two-dimensional electronic spectroscopy (2DES) has also been
applied to both isolated plasmonic and plexcitonic
systems to investigate coherent interactions and population
dynamics, revealing the time scales of various energy
relaxation pathways \cite{finkelstein21,peruffo23}.

Theoretical models and simulations are essential to
understand the energy pathways and the spectroscopy signals
observed in experiments.\cite{gonzalez13,zengin13,zengin15,
santhosh16,leng18a,pelton19,finkelstein21}.
Based on the classical model of coupled harmonic 
oscillators,\cite{rudin99} Faucheaux  {\it et al.} 
investigated the transition in scattering line shape 
from Rabi splitting to Fano interference by increasing the 
damping rate of the plasmonic oscillator.\cite{faucheaux14} 
Quantum models have also been developed.
Manjavacas {\it et al.} simulated the
absorption spectra of a two level system coupled to a
nanoparticle plasmon dimer using Zubarev's Green 
functions.\cite{manjavacas11}
Tomasz and coworkers
modeled the photoabsorption spectra of metal nanoparticle
dimers coupled to organic conjugated molecules using 
first-principle calculations.\cite{rossi19,kuisma22} 

Non-Hermitian effective Hamiltonian approaches,
\cite{bender07,dzsotjan16,varguet19} 
which include the damping of both the plasmon and exciton
while considering their coupling, have become increasingly 
popular due to their clear physical picture and low 
computational cost compared to full quantum simulations. 
\citeauthor{varguet19} used this approach to investigate a
two-level system coupled to a multi-mode localized surface
plasmon, revealing a new hybrid energy diagram
and Rabi splitting in the polarization spectrum
\cite{varguet19}. 
Finkelstein-Shapiro {\it et al.} modeled
the linear absorption and 2DES of plexcitons, providing
insights into the time scales of different relaxation
pathways \cite{finkelstein21}. In a subsequent work, they
demonstrated how the linear spectra and shape symmetries in
2DES change across weak and strong coupling
conditions.\cite{finkelstein23}

In this work, we present a theoretical model of a plexciton
system consisting of a molecular exciton coupled to a single
plasmonic mode. The exciton is described as a two-level
system (TLS), while the plasmonic mode is modeled as a
dissipative harmonic oscillator. The TLS can also be coupled
to intramolecular vibrational modes to account for
electronic-vibrational coupling in the exciton. The dynamics
and spectroscopic signals of this plexciton system are then
simulated using the hierarchical equations of motion (HEOM)
method.\cite{tanimura89,tanimura20,yan21} 
The absorption spectra are calculated by deriving the dipole
autocorrelation functions, and the effects of
exciton-plasmon coupling strength and intramolecular
vibrational modes are analyzed. To investigate the nonlinear
response of the plexciton system, two-dimensional electronic
spectra (2DES) are simulated through direct calculations of
the system’s response to external laser pulses.\cite{gelin05}. 
The 2DES for different coupling
strengths are presented, along with their relationship to
field-driven exciton dynamics.

The remaining sections of this paper are organized as
follows. In Sec. \ref{theory}, we introduce the model
Hamiltonian for the plexciton system, provide a brief
overview of the HEOM method, and describe the approaches
used to calculate the linear absorption spectra and 2DES. In
Sec. \ref{result}, we present the results for the absorption
spectra and 2DES under various conditions. Finally,
conclusions and discussions are made in Sec.
\ref{discussions}.

\section{Theory}
\label{theory}
\subsection{Model Hamiltonian}
We consider the molecular exciton as a TLS, and describe the 
plasmonic mode as a group of harmonic oscillator modes, which 
is also equivalent to a single harmonic oscillator mode
coupled to a dissipative bath.\cite{garg85}
The Hamiltonian of the total system is given by:
\begin{equation}
\label{eq-H_1}
   H = H_e + H_{p} + H_{e-p} \;\;.
\end{equation}
$H_e$ describes the electronic degrees of freedom (DOFs) 
of the exciton with
\begin{equation}
    H_e = \left( \begin{array}{cc}
    0   & 0 \\
    0   & \omega_0
    \end{array} \right),  
\end{equation}
where $\omega_0$ is the transition energy of the TLS.
$H_{p}$ is the Hamiltonian of harmonic oscillator bath used to 
describe the plasmon mode, and is given by
\begin{equation}
    H_{p} = \sum_{j=1}^{N_b} \left(\frac{p_j^2}{2} + 
    \frac{1}{2}\omega_j^2 x_j^2\right) \;\;,
\end{equation}
where the $N_b$ is the number of the bath modes,
$x_j$ and $p_j$ are the position and momentum of the $j$th harmonic 
oscillator bath mode with frequency $\omega_j$.

The exciton-plasmon coupling $H_{e-p}$ is assumed to arise from 
dipole-dipole coupling between the TLS and the plasmon,
\begin{equation}
  H_{e-p} = - \alpha\sum_{j=1}^{N_b} c_j x_j \otimes \sigma_x
    = -\alpha F \otimes \sigma_x \;\;,
\end{equation}
where the collective 
bath coordinate $F = \sum_{j=1}^{N_b}c_j x_j$ 
describes the dipole moment operator of the plasmon modes,
$\sigma_x = |0 \rangle \langle 1| + |1 \rangle \langle 0|$ 
is the Pauli matrix associated with the transition dipole 
of the TLS, $\alpha$ is a dimensionless coefficient.
The spectral density $J(\omega)$ is defined as:
\begin{equation}
    J(\omega) = \frac{\pi}{2} \sum_{j=1}^{N_b} \frac{c_j^2}{\omega_j}
    \delta(\omega - \omega_j) \;\;.
\end{equation}
The autocorrelation function of the collective bath
coordinates $F$ is then given by:
\begin{equation}
\label{eq-bathcf}
    C(t) = \frac{1}{Z_B} \mathrm{Tr} 
    (e^{-\beta H_{p}} e^{iH_{p}t} F e^{-iH_{p}t} F)
    = \frac{1}{\pi} \int_{-\infty}^{\infty} d\omega 
    \frac{e^{-i\omega t}J(\omega)}{1-e^{-\beta \omega}} \;\;,
\end{equation}
where we have extended the definition of the spectral density to 
$\omega <0 $, with $J(-\omega) = - J(\omega)$.

In this study, we employ a Lorentzian type spectral density
to describe the damped harmonic
oscillator:\cite{garg85,tanaka09,li22} 
\begin{equation}
    \label{eq-lorentz1}
    J(\omega) = 4\lambda_0 \frac{\gamma \Omega^2 \omega}
    {(\Omega^2 - \omega^2)^2 + 4\gamma^2 \omega^2} \;\;,
\end{equation}
or its equivalent form:\cite{meier99,song15,li22}
\begin{equation}
    \label{eq-lorentz2}
    J(\omega) = \frac{p_0 \omega}{[(\omega + \omega_p)^2 + \gamma^2]
    [(\omega - \omega_p)^2 + \gamma^2]} \;\;,
\end{equation}
with the relationships of $\Omega^2 = \omega_p^2 + \gamma^2$
and $p_0 = 4 \lambda_0 \Omega^2 \gamma$.
Here, $\gamma^{-1}$ describes relaxation time of the 
under damped plasmonic mode, $\omega_p$ is the oscillating 
frequency, and $\lambda_0$ is the renormalization energy.

The dipole moment operator consists of contributions from both 
the TLS and the collective bath mode for the plasmons, 
and is given by:
\begin{equation}
\label{eq-mu}
\mu = \mu_{\rm TLS} \sigma_x + \frac{\mu_{\rm P}}{\lambda_0} F \;\;,
\end{equation}
where $\mu_{\rm TLS}$ and $\mu_{\rm P}$ are constants related to the 
transition dipole moment of the TLS and plasmon mode.
Since both $F$ and $\lambda_0$ have units of energy, $\mu_{\rm TLS}$ 
and $\mu_{\rm P}$ also have the same units.

\subsection{HEOM}
In the HEOM formalism, the bath correlation function in
Eq.(\ref{eq-bathcf}) is first written into a sum of
exponential decaying functions in time:
\begin{equation}
\label{eq-cf}
    C(t) = \sum_{k=0}^{K-1} c_k e^{-\gamma_k t} \;\;.
\end{equation}
In this work, the barycentric spectral decomposition (BSD)
method\cite{xu22,dan23a} is used to obtain the exponential
decomposition of the bath correlation function. This
approach has been demonstrated to handle general forms of
spectral densities and low-temperature conditions with high
accuracy and efficiency.\cite{xu22,dan23a}

The BSD methods works in the frequency domain. The product
of the spectral density $J(\omega)$ and the Bose function
$1/(1-e^{-\beta\omega})$ in Eq. (\ref{eq-bathcf}) is first
fitted by a summation of fractions:
\begin{equation}
    \frac{J(\omega)}{1-e^{-\beta\omega}} = 
    \sum_{k}\frac{r_k}{\omega - p_k} \;\;.
\end{equation}
The residuals $\{ r_k\}$ and poles $\{ p_k \}$ are then used for the 
exponential decomposition of the bath correlation function:
\begin{equation}
\begin{aligned}
\label{eq-cf-exp2}
 C(t) &= \frac{1}{\pi} \int_{-\infty}^{\infty} d\omega \,
  \frac{e^{-i\omega t}J(\omega)}{1-e^{-\beta \omega}} 
  = -2i\sum_k r_k e^{-ip_kt} \;\;.
\end{aligned}
\end{equation}
It is noticed that, the Cauchy contour integral in Eq. (\ref{eq-cf-exp2})
requires only the poles $p_k$s with a negative
imaginary part in the summation.
The complex coefficients and decaying constants in Eq.(\ref{eq-cf}) 
are thus given by:
\begin{equation}
    c_k = -2i r_k, \;\; \gamma_k = i p_k \;\;.
\end{equation}

By applying the dynamics filtering approach with the renormalized 
auxiliary density operators (ADOs),\cite{shi09c}
\begin{equation}
\label{eq:scale-rho}
    \widetilde{\rho}_{\mathbf{m,n}} = 
    \frac{\rho_{\mathbf{m,n}}}{\prod_k \sqrt{|c_k|^{m_k} m_k!}
    \sqrt{|c_k^*|^{n_k} n_k!}} \;\;,
\end{equation}
the formal exact HEOM are then given by:
\begin{align}
\label{eq-HEOM}
    \frac{d}{dt} \widetilde{\rho}_{\bf m,n} = &-\left(i\mathcal{L} 
    +\sum_k m_k \gamma_k + \sum_k n_k \gamma_k^* \right) 
    \widetilde{\rho}_{\bf m,n} 
    + i \sum_k \sqrt{(m_k+1) c_k} \left[\alpha \sigma_x, 
    \widetilde{\rho}_{{\bf m}_k^+,{\bf n}}\right] \nonumber \\
    & + i \sum_k \sqrt{(n_k+1) c_k^*} \left[\alpha \sigma_x, 
    \widetilde{\rho}_{{\bf m},{\bf n}_k^+}\right] 
    +i \sum_k \sqrt{m_k c_k} \alpha \sigma_x \widetilde{\rho}_{{\bf m}_k^-,{\bf n}} 
    +i\sum_k\sqrt{n_k c_k^*} 
    \widetilde{\rho}_{{\bf m},{\bf n}_k^-}\alpha \sigma_x \;\;.
\end{align}
where $\mathcal{L}\widetilde{\rho} = [H_e,\widetilde{\rho}]$ 
is the Liouville operator. 
The subscripts $\mathbf{n}$ denotes the set of
indices $\mathbf{n} = \{n_{1},n_{2},\cdots\}$, 
and $\mathbf{n}^{\pm}_{k}$ differs from $\mathbf{n}$ only by changing 
the specified $n_{k}$ to $n_{k}\pm1$.
The $\rho_{\bf 0,0}$ term with $\mathbf{0} = \{0,0,\cdots\}$ 
is the system reduced density operator (RDO), while the other 
$\rho_{\mathbf{m,n}}$ terms are the ADOs. 
The ADOs contain important information on the system-bath correlations,
\cite{zhu12,liu14,zhang16b,chen21b}
which will be shown later in our calculations of the dipole
autocorrelation functions and the field-induced polarization.

\subsection{Linear Absorption Spectra}
The linear absorption spectra of the plexciton system are 
obtained by the Fourier transform of 
dipole autocorrelation function:\cite{mukamel95}
\begin{equation}
    I(\omega) \propto \frac{1}{2\pi} \int_{-\infty}^{\infty} dt
    e^{i\omega t} \langle \mu(t)\mu(0) \rangle \;\;,
\end{equation}
where
\begin{equation}
\langle \mu(t)\mu(0) \rangle = \mathrm{Tr} [ e^{i H t}\mu e^{-i H t} 
\mu \rho_{eq}]
= \mathrm{Tr} [\mu e^{-i H t} \mu \rho_{eq} e^{i H t}] \;\;,
\end{equation}
where $\rho_{eq}$ is the equilibrium density operator  
$\rho_{eq} =  e^{-\beta H}/{\mathrm{Tr} e^{-\beta H}}$
($\beta = 1/{k_B T}$ is the inverse temperature).

The calculation process differs slightly from our previous
work on molecular systems\cite{chen09b,jing13} due to the dipole operator
defined in Eq. (\ref{eq-mu}), which includes not only the
system (TLS) part but also contributions from the bath. As a
result, the initial state used for real-time propagation
with the HEOM is given by:\cite{zhu12,zhang16b,chen21b}
\begin{equation}
\label{eq-initial_rho}
    \rho_{\mathbf{m,n}}(t=0,\mu) = 
    \mu_{\rm TLS} \sigma_x \rho^{eq}_{{\bf m},{\bf n}}
    + \frac{\mu_{\rm P}}{\lambda_0} \sum_k 
    \left(\rho^{eq}_{{\bf m}_{k}^{+},{\bf n}}
    +\rho^{eq}_{{\bf m},{\bf n}_{k}^{+}}
    + m_k c_k \rho^{eq}_{{\bf m}_{k}^{-},{\bf n}} \right)\;\;,
\end{equation}
and after renormalization in Eq.(\ref{eq:scale-rho}), it becomes
\begin{equation}
    \widetilde{\rho}_{\mathbf{m,n}}(t=0,\mu) = 
    \mu_{\rm TLS} \sigma_x \widetilde{\rho}^{eq}_{{\bf m},{\bf n}}
    - \frac{\mu_{\rm P}}{\lambda_0} \sum_k 
    \left(\sqrt{(m_k + 1) c_k} \widetilde{\rho}^{eq}_{{\bf m}_{k}^{+},{\bf n}}
    +\sqrt{(n_k + 1) c_k^*}\widetilde{\rho}^{eq}_{{\bf m},{\bf n}_{k}^{+}}
    + \sqrt{m_k c_k} \widetilde{\rho}^{eq}_{{\bf m}_{k}^{-},{\bf n}} \right)\;\;,
\end{equation}
The minus sign in the above equation is due to the convention in defining
the ADOs.\cite{zhu12}

After obtaining the RDO/ADOs $\rho_{\mathbf{m,n}}(t,\mu)$ at
time $t$ through the propagation of the HEOM, we can
calculate the auto-dipole correlation as: \cite{zhang16b,chen21b}
\begin{equation}
\label{eq-corr}
\begin{aligned}
    \langle \mu(t)\mu(0) \rangle
    & =\mu_{\rm TLS} \sigma_x \mathrm{Tr}\left[\rho_{\mathbf{0,0}}(t,\mu)\right]
    - \frac{\mu_{\rm P}}{\lambda_0} \mathrm{Tr} \left[ \sum_k \left[
        \rho^{k}_{\mathbf{1,0}}(t,\mu) + 
        \rho^{k}_{\mathbf{0,1}}(t,\mu)\right]\right] \\
    & =\mu_{\rm TLS} \sigma_x \mathrm{Tr}\left[\widetilde{\rho}_{\mathbf{0,0}}(t,\mu)\right]
    - \frac{\mu_{\rm P}}{\lambda_0} \mathrm{Tr} 
    \left[ \sum_k \left[
        \sqrt{c_k} \widetilde{\rho}^{k}_{\mathbf{1,0}}(t,\mu) + 
        \sqrt{c_k^*} \widetilde{\rho}^{k}_{\mathbf{0,1}}(t,\mu)\right]\right],
\end{aligned}
\end{equation}
where $\rho_{\mathbf{0,0}}(t,\mu)$ is the RDO, $\rho_{\mathbf{1,0}}(t,\mu)$s
and $\rho_{\mathbf{0,1}}(t,\mu)$s are the first-tier ADOs.

\subsection{2DES}
\label{method:2des}
When calculating 2DES, we employ the previously developed
method to directly calculate molecular responses with
explicit laser fields,\cite{seidner95,gelin09,leng17} which
naturally takes account into the effects of finite laser pulse
durations. To this end, the Hamiltonian describing the
interaction between the laser field and the plexciton system
is defined as:
\begin{equation}
\label{eq-H_int}
 H_{int} = -\mu E(t) = -\mu_{\rm TLS} E(t) \sigma_x 
 - \frac{\mu_{\rm P}}{\lambda_0} E(t) F \;\;,
\end{equation}
where $\mu$ represents the total dipole operator of the
plexciton system. The laser pulses are modeled as classical
electromagnetic fields, with the electric field expressed
as:
\begin{equation}
  E(t) = \sum_{i=1}^{3} E_i(t) e^{-i(\omega_i t- {\bf k}_i\cdot {\bf r})} + c.c. \;\;,
\end{equation}
where the summation accounts for the three laser pulses used
in the 2DES experiment. 

The total Hamiltonian, obtained by combining Eqs. (\ref{eq-H_1}) 
and (\ref{eq-H_int}), is given by:
\begin{equation}
    H_T(t) = \left[H_e - \mu_{\rm TLS} \sigma_x E(t)\right] + H_{p} 
    - \left[\alpha\sigma_x + 
    \frac{\mu_{\rm P}}{\lambda_0} E(t)\right] \otimes F \;\;.
\end{equation}
The above Hamiltonian contains the interaction of the
external field and the bath DOFs, and the corresponding HEOM
have been obtained previously in Refs.\cite{zhang16b,wang18a}:
\begin{align}
\label{eq-HEOM-field}
    \frac{d}{dt} \widetilde{\rho}_{\bf m,n} = &-\left(i\mathcal{L}(t) 
    +\sum_k m_k \gamma_k + \sum_k n_k \gamma_k^* \right) 
    \widetilde{\rho}_{\bf m,n} 
    + i \sum_k \sqrt{(m_k+1) c_k} 
    \left[\alpha \sigma_x+\frac{\mu_{\rm P}}{\lambda_0} E(t), 
    \widetilde{\rho}_{{\bf m}_k^+,{\bf n}}\right] \nonumber \\
    & + i \sum_k \sqrt{(n_k+1) c_k^*} 
    \left[\alpha \sigma_x+\frac{\mu_{\rm P}}{\lambda_0} E(t), 
    \widetilde{\rho}_{{\bf m},{\bf n}_k^+}\right] 
    +i \sum_k \sqrt{m_k c_k} \left[\alpha \sigma_x+\frac{\mu_{\rm P}}{\lambda_0} E(t)\right]
    \widetilde{\rho}_{{\bf m}_k^-,{\bf n}} \nonumber \\
    & + i\sum_k\sqrt{n_k c_k^*} \widetilde{\rho}_{{\bf m},{\bf n}_k^-} 
    \left[\alpha \sigma_x+\frac{\mu_{\rm P}}{\lambda_0} E(t)\right] 
    \;\;.
\end{align}
where $\mathcal{L}(t)\rho = [H_e - \mu_{\rm TLS} \sigma_x E(t), \rho]$ is the 
time-dependent Liouville operator.

The total optical polarization induced by the laser pulses
is given by:
\begin{equation}
    P(t) = \langle \mu\rho(t) \rangle  \;\;,
\end{equation}
where the average is taken over all DOFs. This total
polarization includes contributions from all orders of
interaction with the laser field along different output
directions. It is thus necessary to separate the third order
rephasing (RP) and non-rephasing (NR) signals from the total
polarization.\cite{seidner95,gelin05,tan08,yan09}.  
For this purpose, we adopt the calculation scheme originally
developed by Gelin {\it et al.}, which is based on 
perturbation theory and employs a combination
of auxiliary density operators.\cite{gelin05} This
approach has been widely applied to simulate third-order
spectroscopy signals, including three-pulse photon echo and
two-dimensional electronic spectroscopy
\cite{gelin05,egorova07,cheng07,leng17,leng18}. 

In this framework, the explicit form of the third-order
polarization is:
\begin{equation}
    P^{(3)}(t) = \langle \mathbf{\mu}
    [\rho_1(t) - \rho_2(t) - \rho_3(t) - \rho_4(t) 
    + \rho_5(t) + \rho_6(t) + \rho_7(t)] \rangle \;\;,
\end{equation}
the seven auxiliary density operators are defined as:\cite{gelin05}
\begin{equation}
\label{eq-seven_ADP}
\begin{aligned}
   & \rho_1(t) = \rho(E_1(t), E_2(t), E_3(t)), \;\;
    \rho_2(t) = \rho(E_1(t), E_2(t), 0),  \\
   & \rho_3(t) = \rho(E_1(t), 0, E_3(t)), \;\;
    \rho_4(t) = \rho(0, E_2(t), E_3(t)), \\
   & \rho_5(t) = \rho(E_1(t), 0, 0), \;\; 
    \rho_6(t) = \rho(0, E_2(t), 0),  \;\;
    \rho_7(t) = \rho(0, 0, E_3(t)) \;\;. 
\end{aligned}
\end{equation}

Whenever there is a zero in the above
definitions, it indicates that the corresponding laser pulse
is omitted in calculating the total electric field. All
auxiliary density operators are computed using the
time-dependent HEOM in Eq. (\ref{eq-HEOM-field}).
In order to obtain the 2D spectra, the RP and NR signals are
extracted using the phase-matching approach (PMA),\cite{hamm11}
where a specific component of the laser field is selectively
incorporated into the simulation. For instance, when
calculating the RP signal along the direction
$\mathbf{k}_{\mathrm{I}} = -\mathbf{k}_1 + \mathbf{k}_2 +
\mathbf{k}_3$, we use
\begin{equation}
\begin{aligned}
    & E_1(t) = E_1 e^{-\frac{(t-t_1)^2}{2\sigma^2}} 
    e^{i[\omega_1 (t - t_1) - {\bf k}_1 \cdot {\bf r}]} , \\
    & E_2(t) = E_2 e^{-\frac{(t-t_2)^2}{2\sigma^2}} 
    e^{-i[\omega_2 (t - t_2) - {\bf k}_2 \cdot {\bf r}]} , \\
    & E_3(t) = E_3 e^{-\frac{(t-t_3)^2}{2\sigma^2}} 
    e^{-i[\omega_3 (t - t_3) - {\bf k}_3 \cdot {\bf r}]} \;\;.  
\end{aligned}
\end{equation}
For the NR signal associated with $\mathbf{k}_{\mathrm{II}} 
= \mathbf{k}_1 - \mathbf{k}_2 +\mathbf{k}_3$, we have:
\begin{equation}
\begin{aligned}
    & E_1(t) = E_1 e^{-\frac{(t-t_1)^2}{2\sigma^2}} 
    e^{-i[\omega_1 (t - t_1) - {\bf k}_1 \cdot {\bf r}]} , \\
    & E_2(t) = E_2 e^{-\frac{(t-t_2)^2}{2\sigma^2}} 
    e^{i[\omega_2 (t - t_2) - {\bf k}_2 \cdot {\bf r}]} , \\
    & E_3(t) = E_3 e^{-\frac{(t-t_3)^2}{2\sigma^2}} 
    e^{-i[\omega_3 (t - t_3) - {\bf k}_3 \cdot {\bf r}]} \;\;.  
\end{aligned}
\end{equation}

Once the third-order responses for the RP and
NR signals are obtained, the 2DES can be calculated by
performing a double Fourier transform with respect to the
coherence time $\tau$ and the detection time $t$ as:
\cite{gelin09,cheng07,cheng08}
\begin{equation}
    S(\omega_{\tau}, T_2, \omega_t) = \mathrm{Re} 
    \int_{0}^{\infty} d\tau \int_{0}^{\infty} dt' 
    [e^{-i \omega_{\tau} \tau + i \omega_t t'}\times i P_{rp} (\tau, T_2, t')
    + e^{i \omega_{\tau} \tau + i \omega_t t'}\times i P_{nr} (\tau, T_2, t')] ,
\end{equation}
where $\omega_{\tau}$ is the excitation frequency,
$\omega_t$ is the detection frequency, $\tau = t_2-t_1$ is the 
coherence time, $T_2 = t_3 -t_2$ is the waiting time, 
$t' = t-t_3$ is the detection time.

\section{Results}
\label{result}

\subsection{Polulation dynamics of the coupled system}

The above HEOM are then applied to a model consisting of a
quantum emitter described by a TLS, and a single dissipative plasmon
mode as the harmonic bath. All the simulations are performed at room
temperature of 298K. The transition energy of the exciton is
$\omega_0 = 15000 \ \mathrm{cm}^{-1}$. We also use $\omega_p
= 15000 \ \mathrm{cm^{-1}}$ in the Lorentzian spectral
density in Eq. (\ref{eq-lorentz1}), assuming energy resonance
between the emitter and plasmon. The other parameters of
Lorentzian spectral density are $\gamma^{-1} = 5 \
\mathrm{fs}$ corresponding to fast decay of the plasmon mode.
\cite{shahbazyan13} The renormalization energy
$\lambda_0$ = 0.05 a.u. (1.1$\times 10^4~\mathrm{cm^{-1}}$). 
By using the equivalence of the TLS-Lorentzian bath model
and the TLS-Harmonic oscillator-Ohmic bath model,\cite{garg85} 
we can calculate the effective coupling between the TLS and the
first excited state of the undamped harmonic plasmon mode, 
which is given by $\Delta = \alpha\sqrt{\lambda_0\Omega}$, where 
$\alpha$, $\lambda_0$, and $\Omega$ are given in Sec. 2.1. 
For $\alpha =0.1$, the coupling constant is 
$\Delta =1287\ \mathrm{cm^{-1}}$,
which satisfies the condition of the strong-coupling regime. 
\cite{novotny10,hummer13}
The transition dipole moments for the TLS and plasmon are 
taken as $\mu_{\rm TLS} = 1$ and $\mu_{\rm P} = 10$, respectively.

The excited-state population dynamics of the TLS are obtained
for various coupling strengths $\alpha$ and are shown in
Fig. \ref{pop-coupling}, where the initial state is a factorized
state with the TLS in the excited state and the plasmon mode
in the thermal equilibrium. For simplicity, coupling of the TLS to
intramolecular vibrational modes is not included. In the case of weak
coupling $\alpha = 0.01$, the population dynamics show a
slow decay with a time constant of approximately $300 \
\mathrm{fs}$. For larger coupling strengths from $\alpha = 0.1$
and $\alpha = 0.2$, the population dynamics exhibit Rabi
oscillations, although with rapid decay caused by the short
lifetime of the plasmon mode.  The $\alpha = 0.05$ case represents an
intermediate case.

\subsection{Linear Absorption spectra}
\begin{figure}[htbp]
    \centering
    \includegraphics[width=14cm]{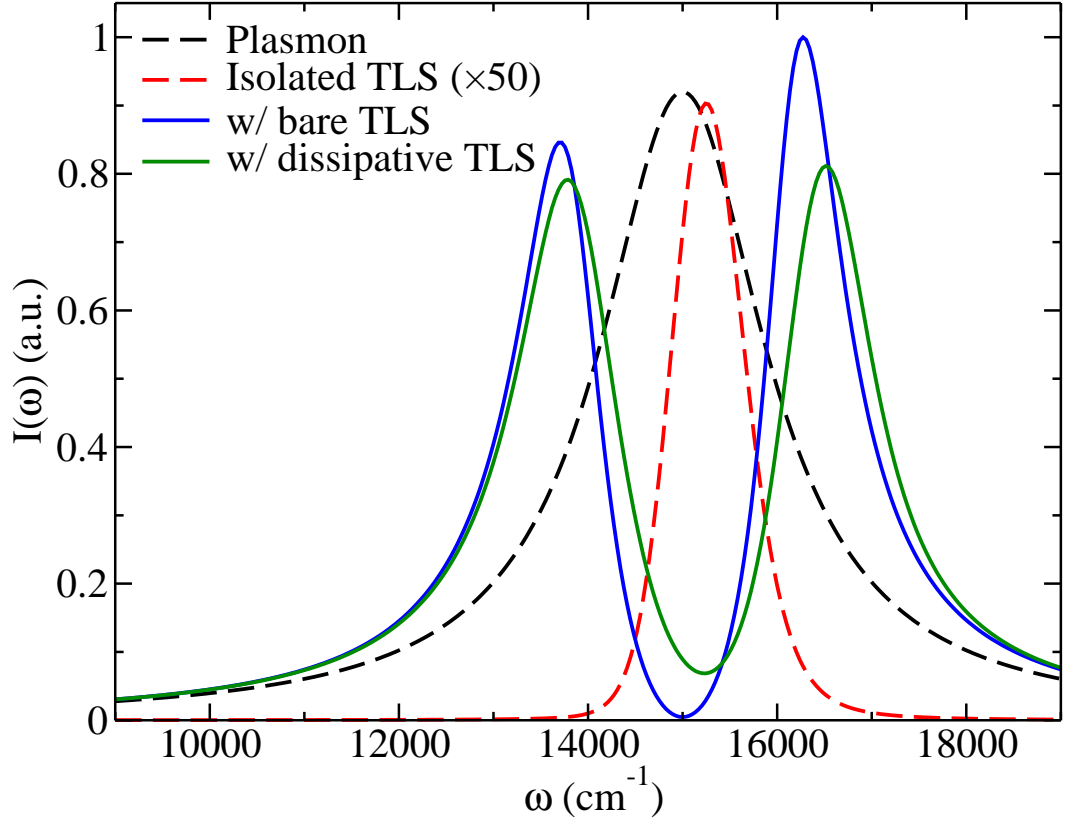}
    \vspace{1em}
    \caption{Linear absorption spectra of the plasmon mode (black dashed line),
    an isolated TLS with coupling to the vibrational DOFs described by 
    a Debye-Drude spectral density (red dashed line), the bare TLS coupled to 
    the plasmon mode (blue solid line), and the TLS coupled simultaneously to the 
    plasmon mode and vibrational DOFs (green solid line). The red line is 
    rescaled by 50 times for better visualization.}
    \label{fig-abs-all}
\end{figure}

The simulated absorption spectra of the TLS-plasmon coupling 
strength $\alpha = 0.1$ are presented in Fig. \ref{fig-abs-all}.
The absorption spectra of the plasmon mode is shown in the 
black dashed line. The broad width of the peak arises from the 
short lifetime of the plasmon mode.
The red dashed line represents the absorption spectra of the isolated TLS.
When there is coupling between the TLS and the plasmon mode,
we consider two cases: one where the TLS is not coupled to
the intramolecular vibrational DOFs, referred to as the
``bare TLS," and another where the TLS is coupled to the
vibrational DOFs that leads to decoherence effects, 
referred to as the ``dissipative TLS".

In the latter case, to take account into effects of coupling 
between intramolecular 
vibrational modes and the electronic transition, we employ a 
shifted harmonic oscillator model to describe the ground and excited 
state potential surfaces.\cite{chen09b,jing13}  In this model,
the coupling between the TLS and the vibrational modes is 
described by a Debye-Drude spectral density:\cite{chen09b,jing13}
\begin{equation}
    J_D (\omega) = \frac{\eta\gamma_{D}\omega}{\gamma_D^2 + \omega^2} \;\;,
\end{equation}
with the parameters $\eta=200 \ \mathrm{cm^{-1}}$ and 
$\gamma_D^{-1}= 40 \ \mathrm{fs}$.
In Fig. \ref{fig-abs-all}, 
the isolated TLS spectra is enlarged by 50 times for better visualization.

The absorption spectra of the plexciton for $\alpha=0.1$,
without and with coupling to vibrational DOFs are shown in 
Fig. \ref{fig-abs-all} as blue and green lines, respectively.
As discussed earlier, this is a case of strong coupling. 
The observed peak splitting is $2723 \ \mathrm{cm^{-1}}$ 
for the case of bare TLS, which is consistent with the period of 
the Rabi oscillations in the population dynamics, as shown in 
Fig. \ref{pop-coupling}. In the spectra with coupling 
to the vibrational modes, the peak splitting is of a similar 
value. 
In the case of the bare TLS (blue line), 
the dip between the two absorption peaks is very deep, 
reaching close to the $\omega$-axis. On the other hand, the 
dip in the presence of TLS-vibrational coupling (green curve) is 
shallower. Additionally, the lower and upper energy peaks are
more symmetric when there is coupling to the vibrational modes.

\begin{figure}[htbp]
    \centering
    \includegraphics[width=12cm]{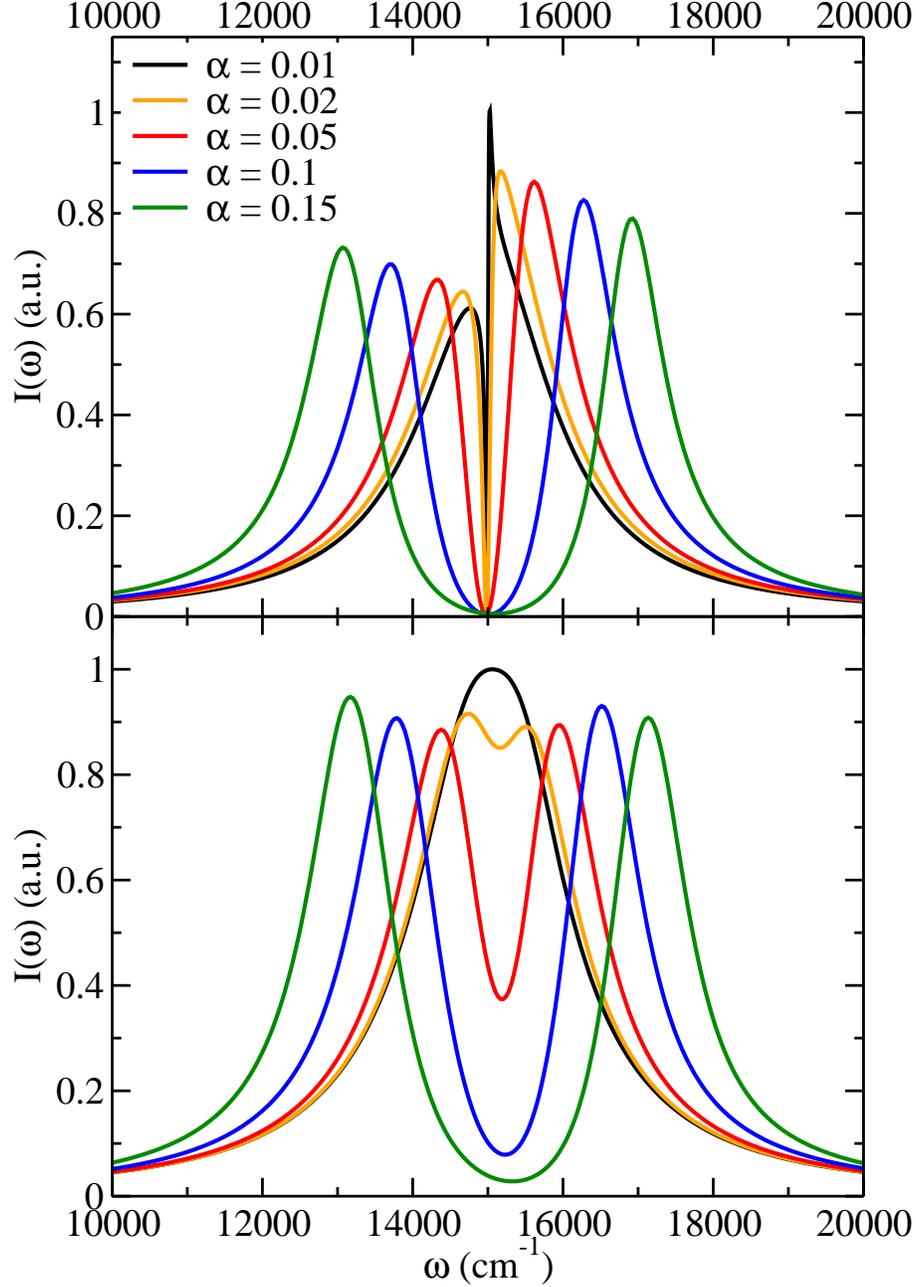}
    \vspace{1em}
    \caption{The absorption spectra of the plexciton system with different
    coupling strength $\alpha$. The upper panel shows the spectra of the bare 
    TLS coupled to the plasmon mode, and the lower panel
    shows results of a TLS coupled simultaneously to the plasmon mode and vibrational DOFs.
    All the other parameters are the same as those in Fig. \ref{fig-abs-all}.}
    \label{fig-abs-coupling}
\end{figure}

The influence of the TLS-plasmon coupling strength 
on the absorption line shape is shown in
Fig. \ref{fig-abs-coupling}, where the upper panel presents the
results for the bare TLS and the lower panel shows results
for the TLS coupled with vibrational modes. 
The dimensionless coupling factor $\alpha$ ranges from weak coupling
($\alpha = 0.01$) to strong coupling ($\alpha = 0.15$).
For $\alpha = 0.01$, the absorption spectrum exhibits a
sharp asymmetric dip, characteristic of the Fano line shape
(black line, upper panel). As the TLS-plasmon coupling
strength increases, the line shape evolves into an
asymmetric double peak. This transition from Fano line shape
to peak splitting caused by Rabi oscillation is consistent 
with previous analysis based on the classical model.\cite{faucheaux14}

The effects of coupling to intramolecular vibrational modes
on the absorption line shape are clearly visible in the
lower panel of Fig. \ref{fig-abs-coupling}. 
In the upper panel, the central dips
are very deep, almost reaching the $\omega$-axis, whereas in
the lower panel, the dips are shallower and vary with the
TLS-plasmon coupling strength. This effect is more pronounced for weak
to intermediate coupling strengths. Notably, in the $\alpha
= 0.01$ case, the dip completely vanishes, resulting in a
single peak that is distinctly different from the line shape
in the upper panel. This likely explains that the Fano line 
shape is rarely observed in the weak coupling regime:
coupling to the vibrational modes is always present and
introduces decoherence to the TLS.\cite{adato13}
For all spectra that retain the double peak structure, the
line width and the amplitude of the peak splitting remain
essentially the same, although the peaks become more symmetric
in the presence of coupling to vibrational modes.

\begin{figure}[htbp]
    \centering
    \includegraphics[width=12cm]{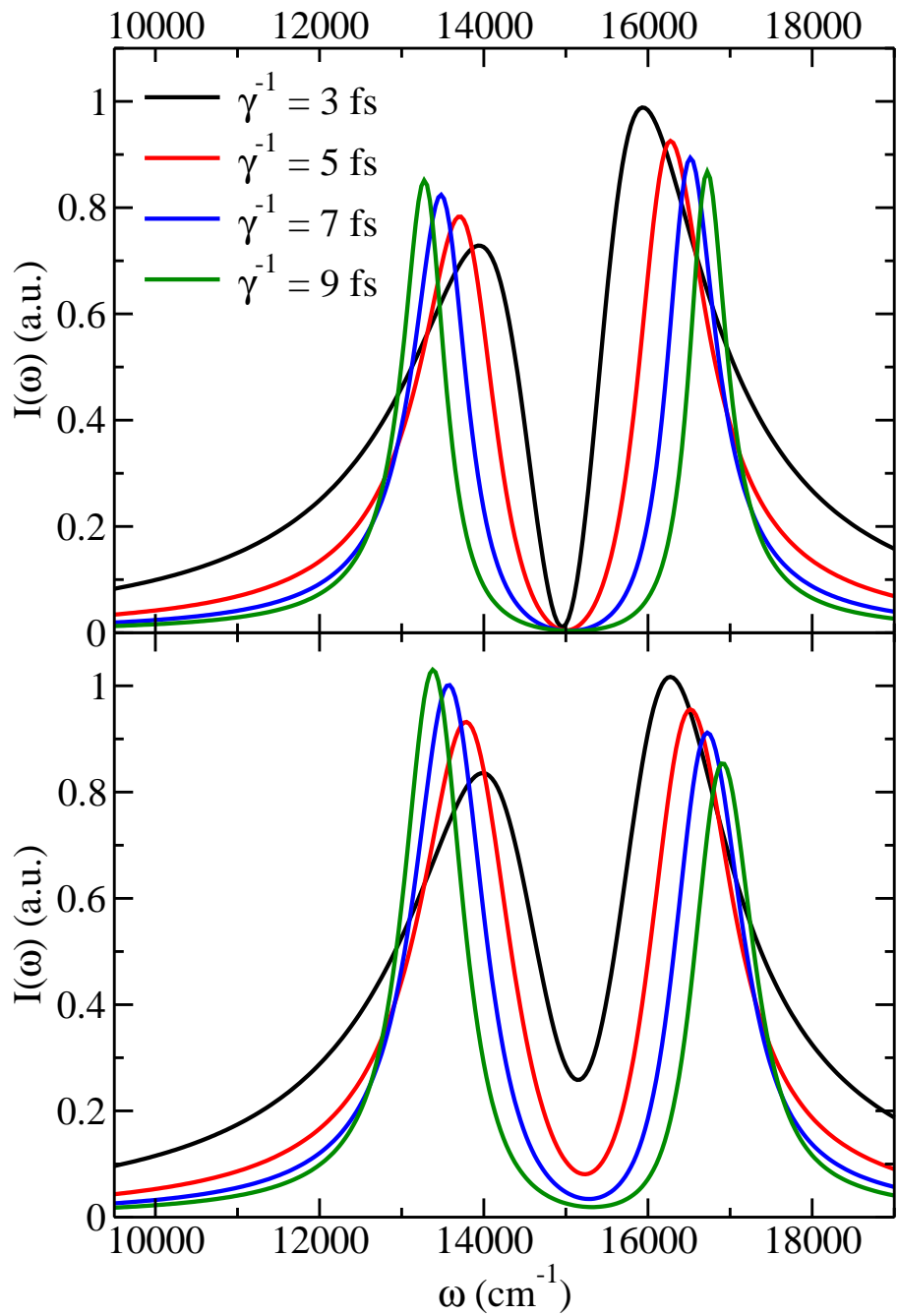}
    \vspace{1em}
    \caption{Same as Fig. \ref{fig-abs-coupling}, for different plasmon mode 
    relaxation rate $\gamma$.
    }
    \label{fig-abs-width}
\end{figure}

Previous theoretical models have suggested that the
transition to Rabi oscillations in a hybrid plexciton system
depends on the damping rate of the plasmon
mode.\cite{faucheaux14,yang16} We thus investigate the
effects of the plasmon damping rate $\gamma$ on the
absorption spectra, with the results shown in Fig.
\ref{fig-abs-width}. All other simulation parameters are
identical to those in Fig. \ref{fig-abs-all} for $\alpha=0.1$.
The most notable effects of $\gamma$ are on the line width
and peak splitting: a smaller plasmon damping rate $\gamma$
results in larger peak splitting and narrower peaks.
A larger damping rate also makes the line shape 
more asymmetric, resembling the Fano line shape
(black curve in the upper panel for $\gamma^{-1} = 3 \
\mathrm{fs}$). These observations are consistent with
previous theoretical
analyses.\cite{manjavacas11,faucheaux14,baranov18}
The effect of coupling to vibrational modes is similar to
the results shown in Fig. \ref{fig-abs-coupling}. It reduces the
depth of the central dip and slightly alters the
ratio of the lower and upper energy peaks, making them more 
symmetric.

\begin{figure}[htbp]
    \centering
    \includegraphics[width=14cm]{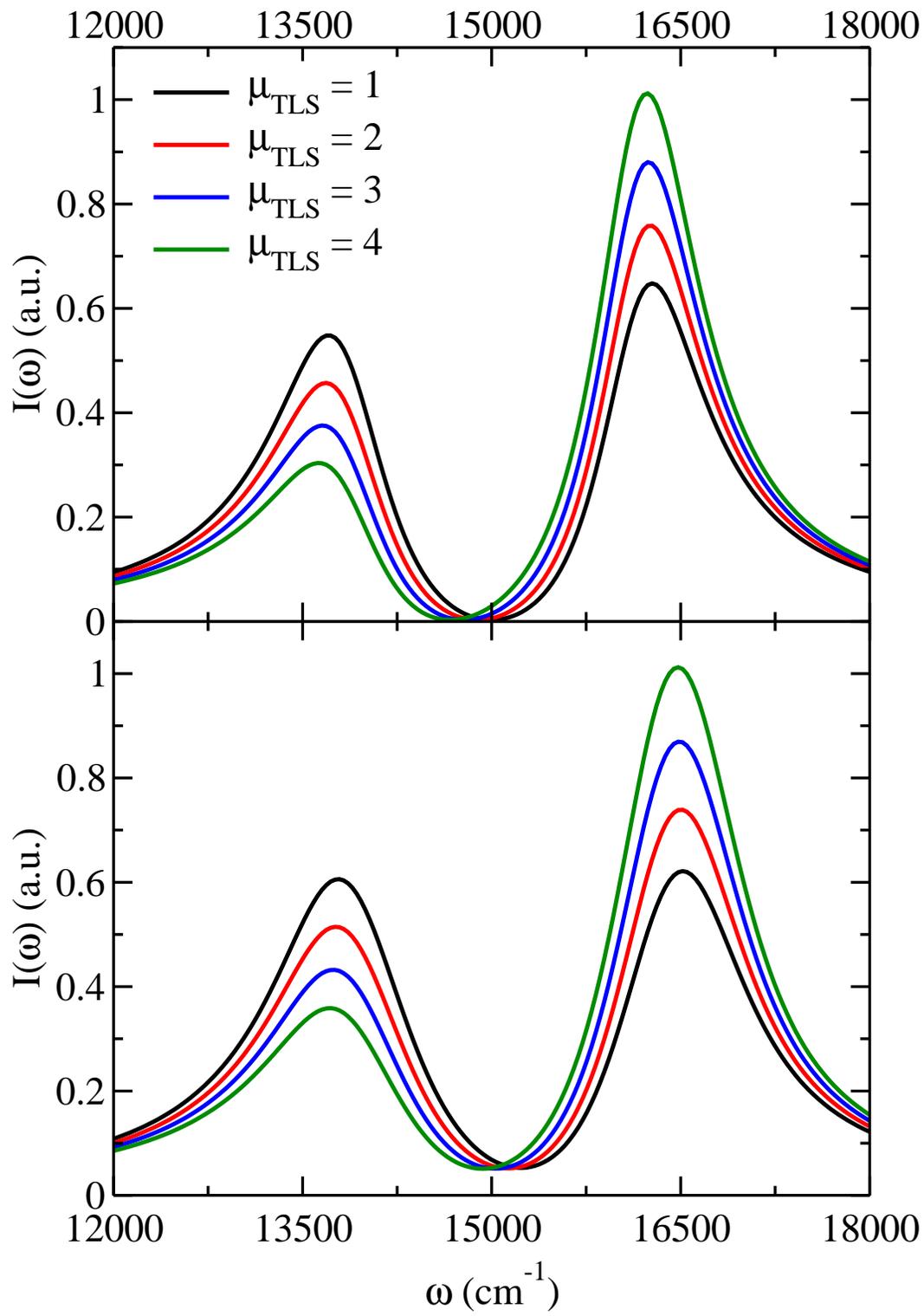}
    \vspace{1em}
    \caption{
    Same as Fig. \ref{fig-abs-coupling}, for different TLS transition dipole moment 
    $\mu_{\rm TLS}$.
    }
    \label{fig-abs-mue}
\end{figure}

Fig. \ref{fig-abs-mue} shows the results obtained by varying the
dipole moment of the TLS, $\mu_{\rm TLS}$, while keeping all other
parameters the same as in Fig. \ref{fig-abs-all}. It is observed
that the ratio of the higher to lower energy peaks increases
as the dipole moment of the TLS increases, and the position
of the central dip redshifts to smaller energies. However,
the energy splitting between the two peaks remains almost
unchanged.
This result is consistent with previous analyses
based on the non-Hermitian Hamiltonian model
that the width of Rabi splitting is only related to the 
coupling strength and damping rates.\cite{baranov18}
Since the simulation parameters correspond to a strong
coupling case ($\alpha = 0.1$), introducing the coupling of
vibrational modes to the TLS (lower panel of Fig.
\ref{fig-abs-mue}) produces effects similar to those in Fig.
\ref{fig-abs-width}.

\subsection{2DES}

\begin{figure}[htbp]
    \centering
    \includegraphics[width=16cm]{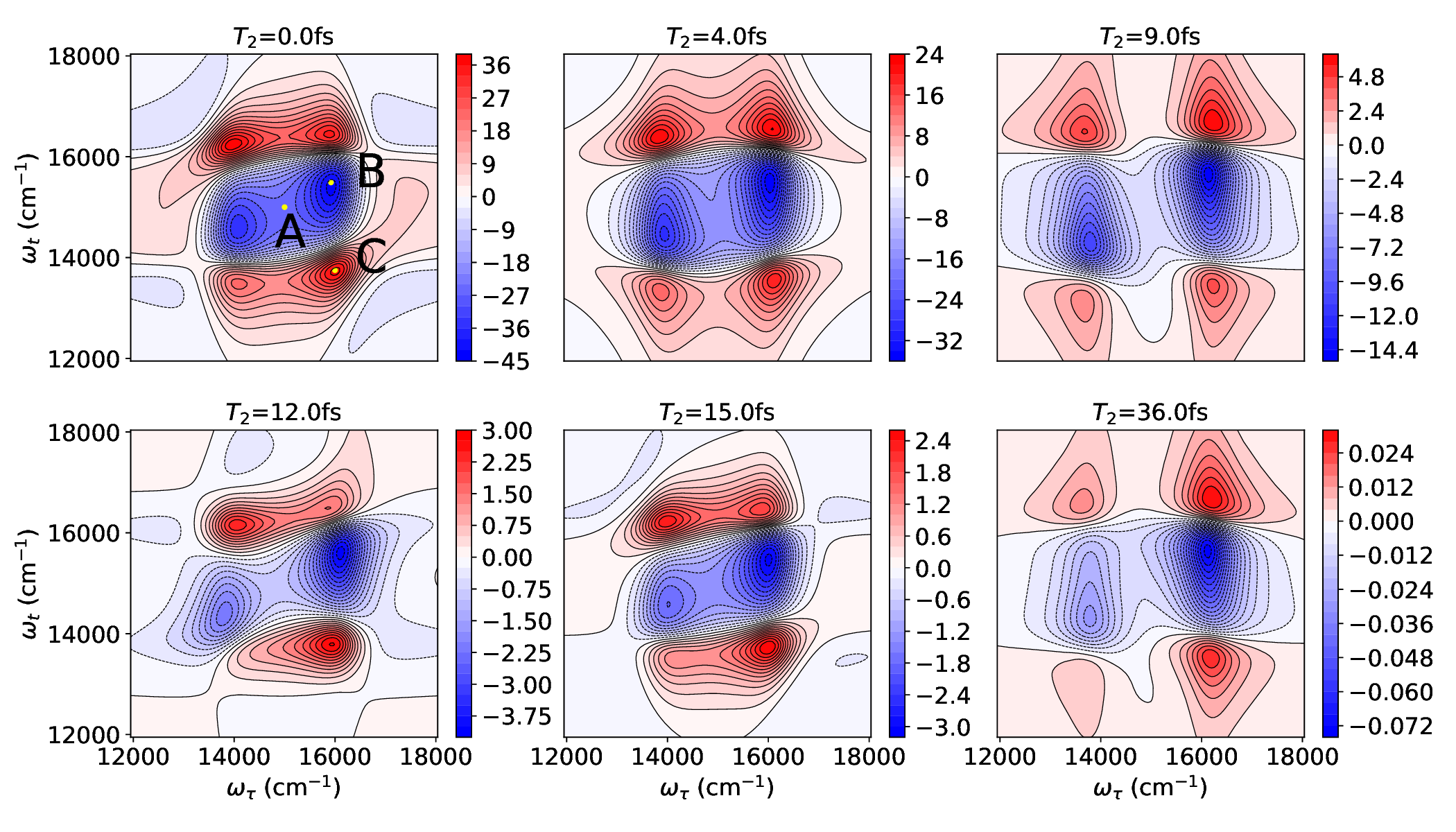}
    \vspace{1em}
    \caption{Simulated 2DES of a bare TLS coupled to the plasmon
    mode at different waiting time $T_2$ for the strong
    coupling case $\alpha = 0.1$. All the other
    parameters are the same as those in Fig.
    \ref{fig-abs-all}.}
    \label{fig-2d-0.1}
\end{figure}

In this subsection, we present results for the 2DES of a TLS
coupled to a plasmon mode with varying coupling strengths.
We first examine the case of $\alpha = 0.1$, which
corresponds to the strong coupling regime. Based on the
previous analysis of the effects of vibrational modes on the
absorption spectra, we note that these effects are not
significant in this regime. We thus focus on the case of a bare
TLS coupling the the plasmon mode, and the results are shown in Fig.
\ref{fig-2d-0.1} for various waiting times, $T_2$. The
calculations are carried out using the method described in
Sec. \ref{method:2des}, where the laser pulse, as defined in
Eq. (\ref{eq-H_int}), has a Gaussian profile centered at
$15000 \ \mathrm{cm}^{-1}$ with a full width at half maximum
(FWHM) of $4 \ \mathrm{fs}$. The spectrum of this 
pulse is sufficiently broad to 
cover the entire absorption spectra of the plexciton system.

\begin{figure}[htbp]
    \centering
    \includegraphics[width=14cm]{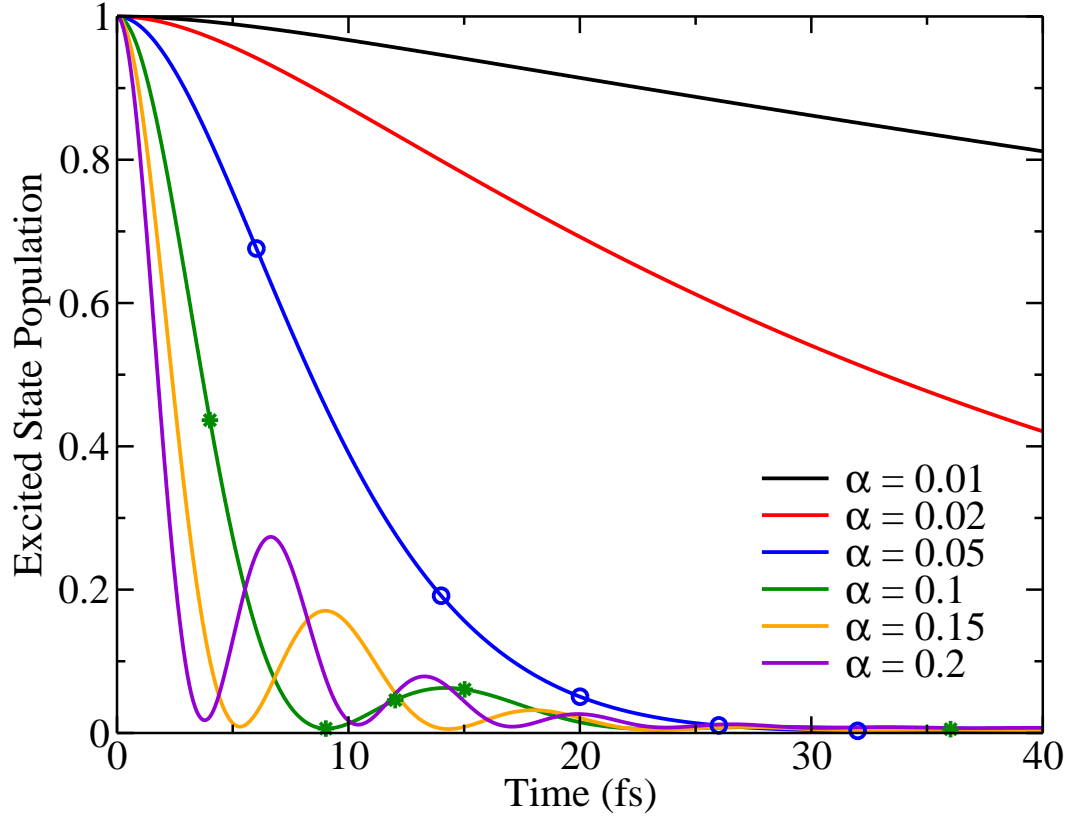}
    \vspace{1em}
    \caption{Excited state population dynamics of the TLS
    with different TLS-plasmon coupling strengths. The 
    initial state is a factorized state with the TLS in the
    excited state and the plasmon mode in the thermal
    equilibrium. Marked symbols on the blue and green curves 
    are used in the later 2DES simulations.}
    \label{pop-coupling}
\end{figure}

To better illustrate the dynamical evolution process, 
we present several representative 2DES at selected
waiting times. These times are chosen based on the
excited-state population dynamics of the TLS as shown in
Fig. \ref{pop-coupling}, and are marked by stars on the 
green curve (population dynamics for $\alpha=0.1$). The
selected waiting times include the initial time at $0 \
\mathrm{fs}$, the first minimum in the population dynamics
at approximately $9 \ \mathrm{fs}$, the peak near $15 \
\mathrm{fs}$, two intermediate times at $4 \ \mathrm{fs}$
and $12 \ \mathrm{fs}$, and a longer waiting time at $36 \
\mathrm{fs}$, when the oscillations have decayed.

The 2DES of the plexciton system exhibit distinct features
compared to those of an isolated TLS, which shows a single
positive peak, or the isolated plasmon mode, where the
nonlinear response of the harmonic system is zero.
It also differs significantly from the 2DES of a dimer
system composed of two coupled
TLSs.\cite{chen10,leng17,gelzinis19} Especially, the diagonal
cut of the 2DES reveals prominent negative peaks, which can
no longer be directly related to the absorption spectra of
the plexciton, as shown in Fig. \ref{fig-abs-all}.

During the Rabi oscillation period from $T_2 = 0$ to 
$15 \ \mathrm{fs}$, the 2DES shows a clear 
``breathing mode" behavior. Along the $\omega_\tau$ axis,
the separation between the two peaks increases from $T_2 = 0
\ \mathrm{fs}$ to $T_2 = 4 \ \mathrm{fs}$, reaches its
maximum at $T_2 = 9 \ \mathrm{fs}$, and then decreases again
to a minimum at $T_2 = 15 \ \mathrm{fs}$. At a longer time,
$T_2 = 36 \ \mathrm{fs}$, the 2DES shows a structure similar
to the maximum separation observed at $T_2 = 9 \
\mathrm{fs}$. However, due to the rapid decay of the plasmon
mode, the amplitude of the 2DES is significantly
smaller compared to that at $T_2 = 0$.

\begin{figure}[htbp]
    \centering
    \includegraphics[width=16cm]{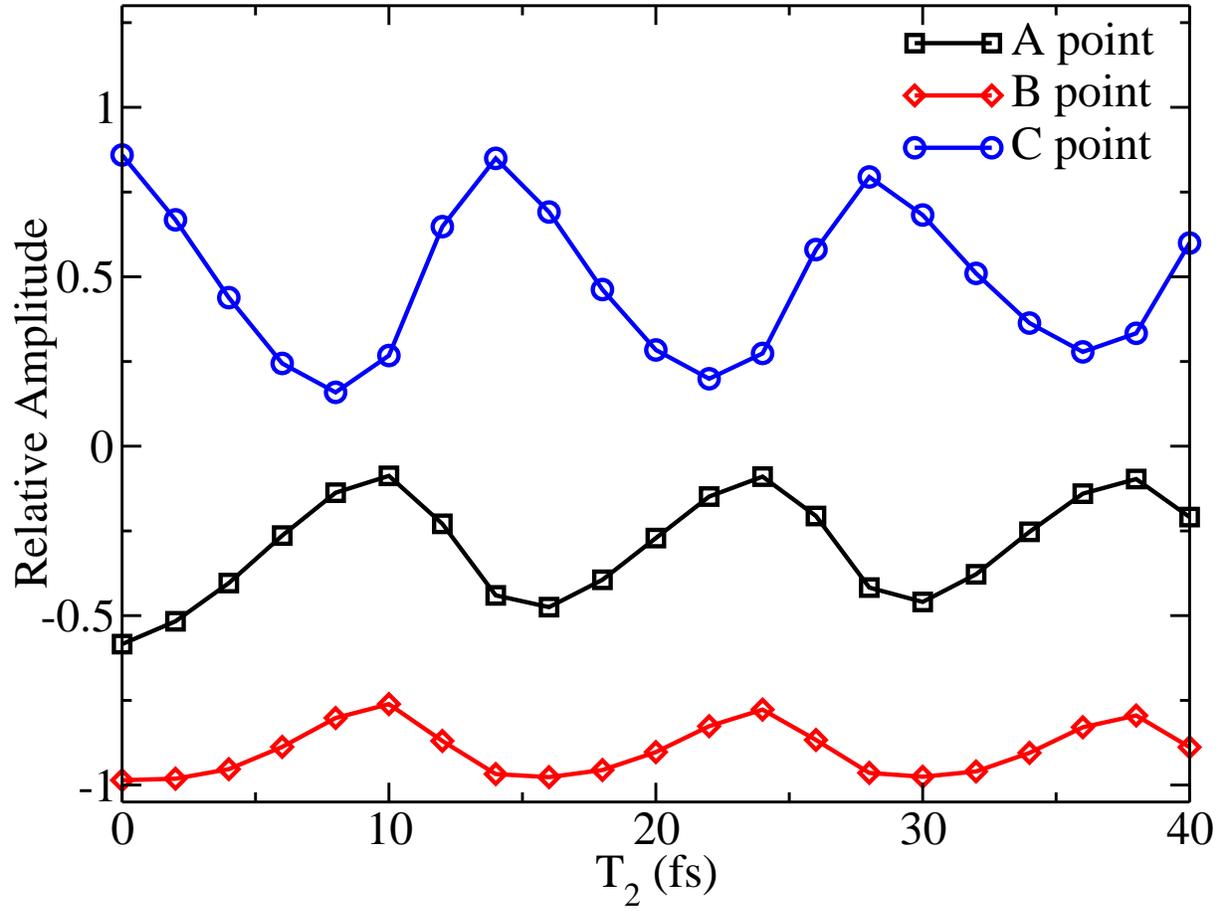}
    \vspace{1em}
    \caption{Time evolution of the relative amplitudes at 
    three selected points in the $\omega_\tau$-$\omega_t$ plane as 
    labeled in Fig. \ref{fig-2d-0.1}. }
    \label{fig-peak-0.1}
\end{figure}

We plot in Fig. \ref{fig-peak-0.1} the evolution of the
amplitudes at three points in the $\omega_\tau$-$\omega_t$
plane, as labeled in Fig. \ref{fig-2d-0.1}: a central point
(A) and two points corresponding to the positive (B) and
negative (C) peaks. To account for the decay of the overall
intensity of the 2DES with increasing $T_2$, the amplitudes
are normalized to the absolute maximum value of each 2DES.
The oscillations of these amplitudes provide further
evidence of the Rabi oscillation.

\begin{figure}[htbp]
    \centering
    \includegraphics[width=16cm]{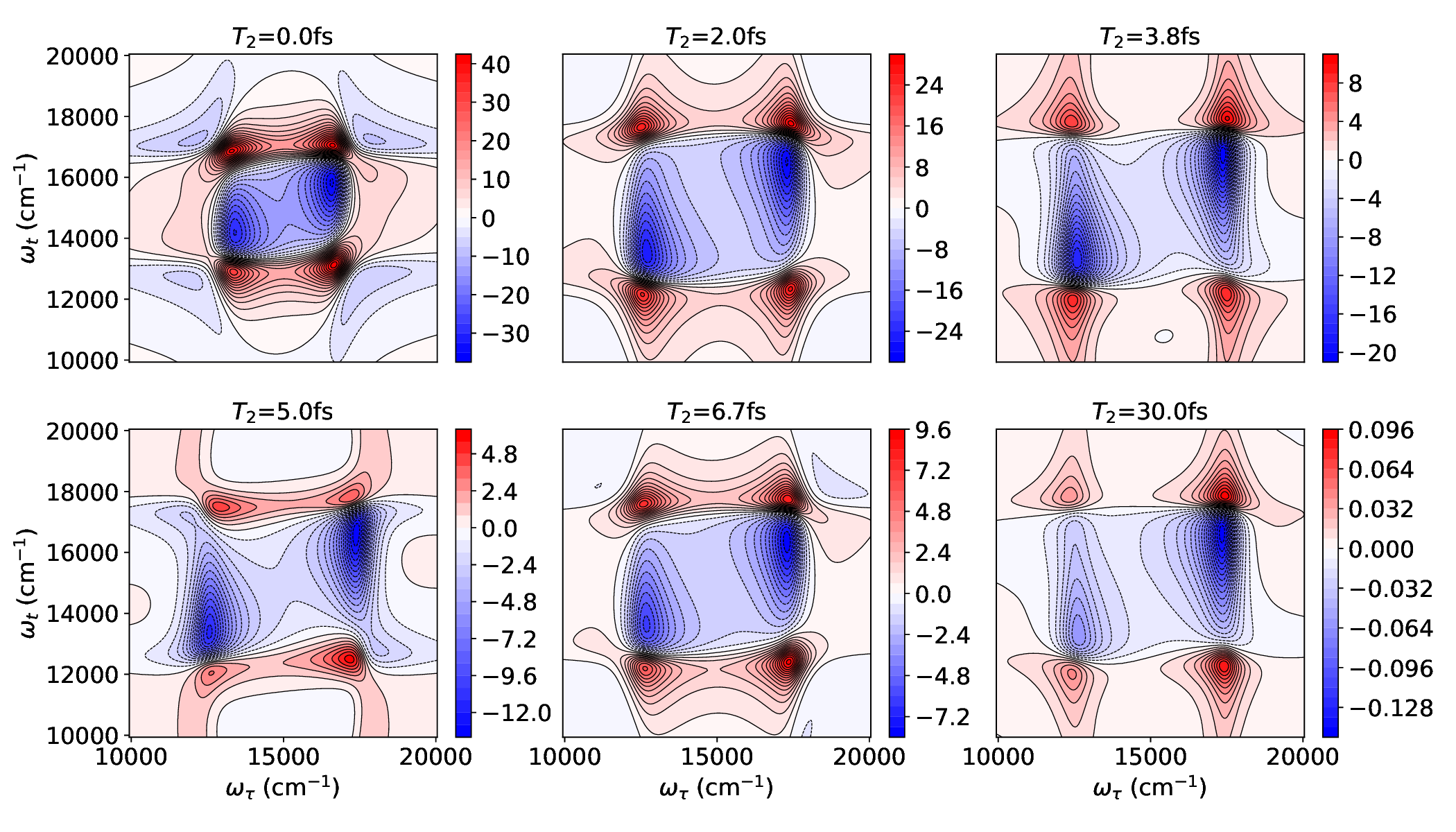}
    \vspace{1em}
    \caption{Same as Fig. \ref{fig-2d-0.1} for $\alpha = 0.2$.
    }
    \vspace{10em}
    \label{fig-2d-0.2}
\end{figure}

We further simulate the 2DES for a larger coupling strength,
$\alpha = 0.2$, with the results presented in Fig.
\ref{fig-2d-0.2}. The patterns show a similar behavior to
those in Fig. \ref{fig-2d-0.1}, but with a significantly
shorter period. The maximum peak separation occurs at $T_2 =
3.8 \ \mathrm{fs}$, followed by a minimum separation at $T_2
= 6.7 \ \mathrm{fs}$.
These results confirm that the ``breathing mode" pattern
observed in the 2DES is indeed associated with the
Rabi oscillations of the plexciton system, as presented in
Fig. \ref{pop-coupling} for the population dynamics of the TLS.

\begin{figure}[htbp]
    \centering
    \includegraphics[width=16cm]{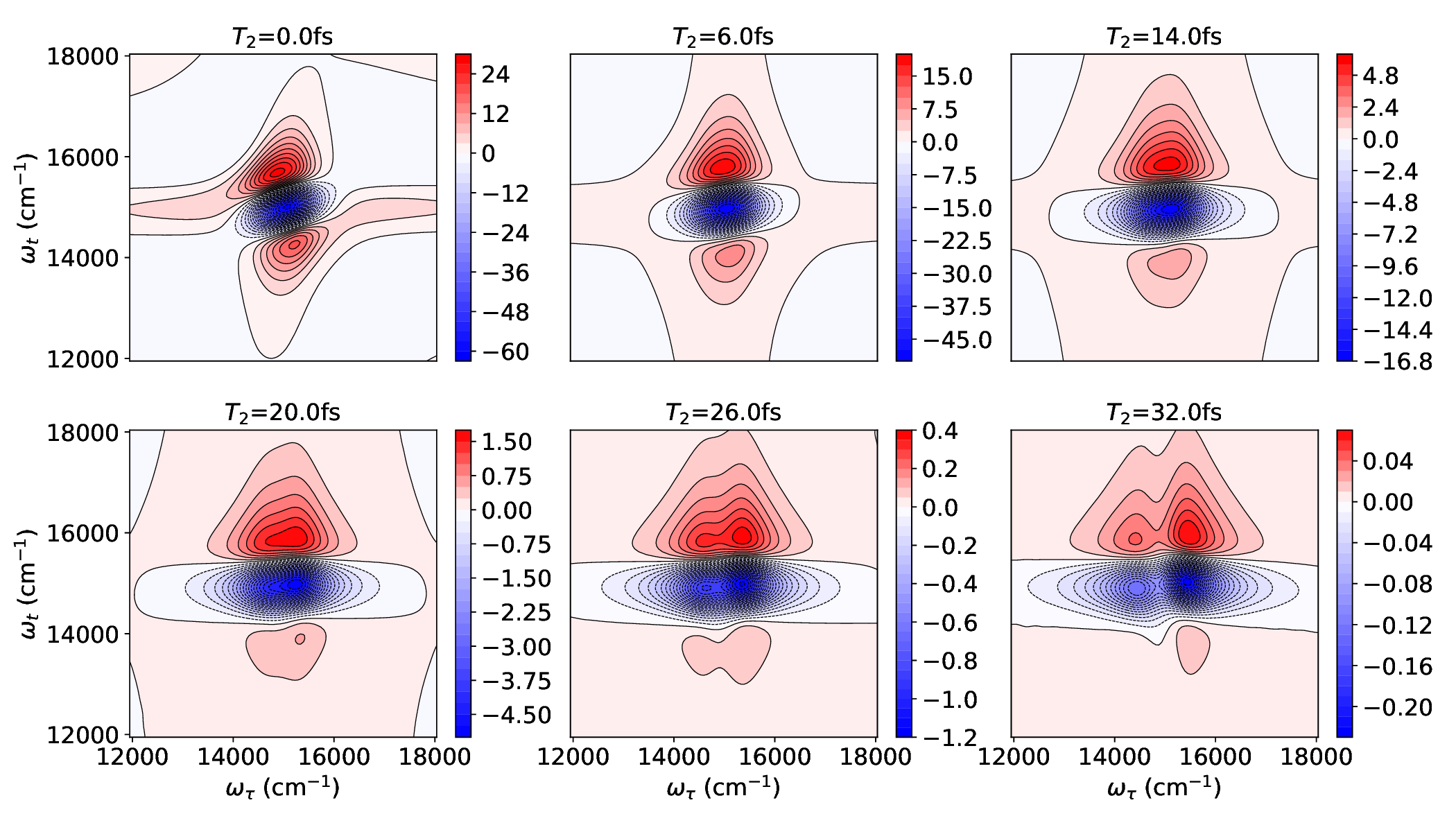}
    \vspace{1em}
    \caption{
    Same as Fig. \ref{fig-2d-0.1} for the intermediate 
    coupling case $\alpha = 0.05$.
    }
    \vspace{10em}
    \label{fig-2d-0.05}
\end{figure}

The 2DES for the intermediate coupling strength, $\alpha =
0.05$, is shown in Fig. \ref{fig-2d-0.05}, with selected
waiting times labeled by blue circles in Fig.
\ref{pop-coupling}. At earlier times, there are single 
positive and negative peaks along the
$\omega_\tau$ direction, even
though the absorption spectra in the intermediate coupling
regime display double peaks as shown in Fig.
\ref{fig-abs-coupling}. This behavior contrasts with the
split peak structures observed in the strong coupling
regime, as seen in Figs. \ref{fig-2d-0.1} and
\ref{fig-2d-0.2}. Furthermore, no ``breathing mode" pattern
is observed, which is consistent with the monotonic decay of
the population dynamics shown in Fig.
\ref{pop-coupling}. At longer times, split peaks appear in
the upper positive and central negative regions at
$T_2 = 26$ and $32 \ \mathrm{fs}$. However, by these waiting
times, the amplitude of the 2DES has decreased
significantly.

\begin{figure}[htbp]
    \centering
    \includegraphics[width=16cm]{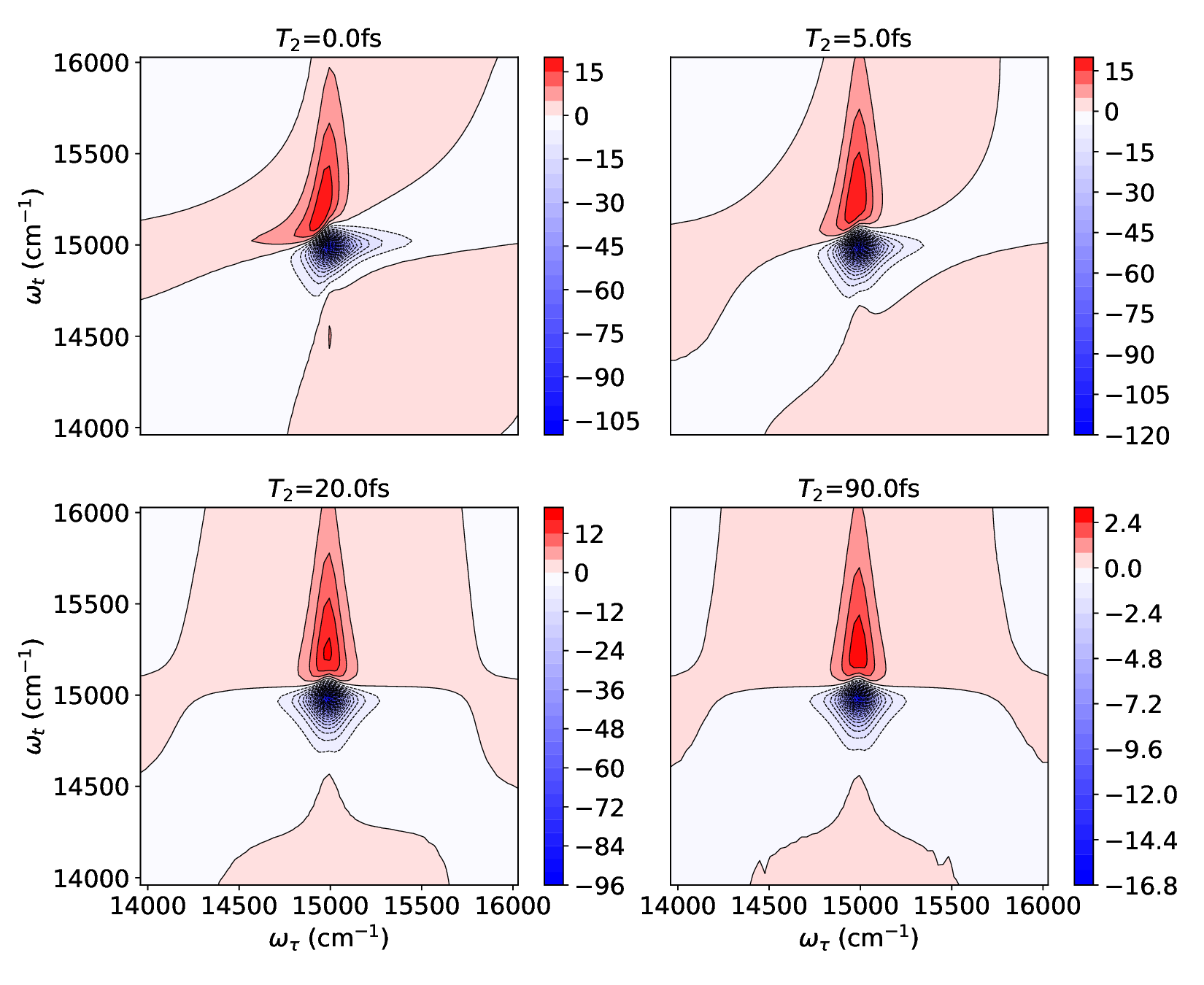}
    \vspace{1em}
    \caption{
    Same as Fig. \ref{fig-2d-0.1} for the weak 
    coupling case $\alpha = 0.02$.
    }
    \label{fig-2d-0.02}
\end{figure}

The 2DES in the weak coupling regime with $\alpha = 0.02$ is
shown in Fig. \ref{fig-2d-0.02}. Compared to the results in
the strong and intermediate coupling regimes, the 2DES
occupies a much smaller area in the $\omega_\tau - \omega_t$
plane and shows only single peaks, with the negative
peak being more prominent.
In addition, the positive peak at smaller
$\omega_t$, which is present in the strong and
intermediate coupling cases, is also absent in Fig.
\ref{fig-2d-0.02}. As the waiting time increases, the
initially asymmetric shapes of the positive and negative
peaks gradually become more symmetric.

\begin{figure}[htbp]
    \centering
    \includegraphics[width=16cm]{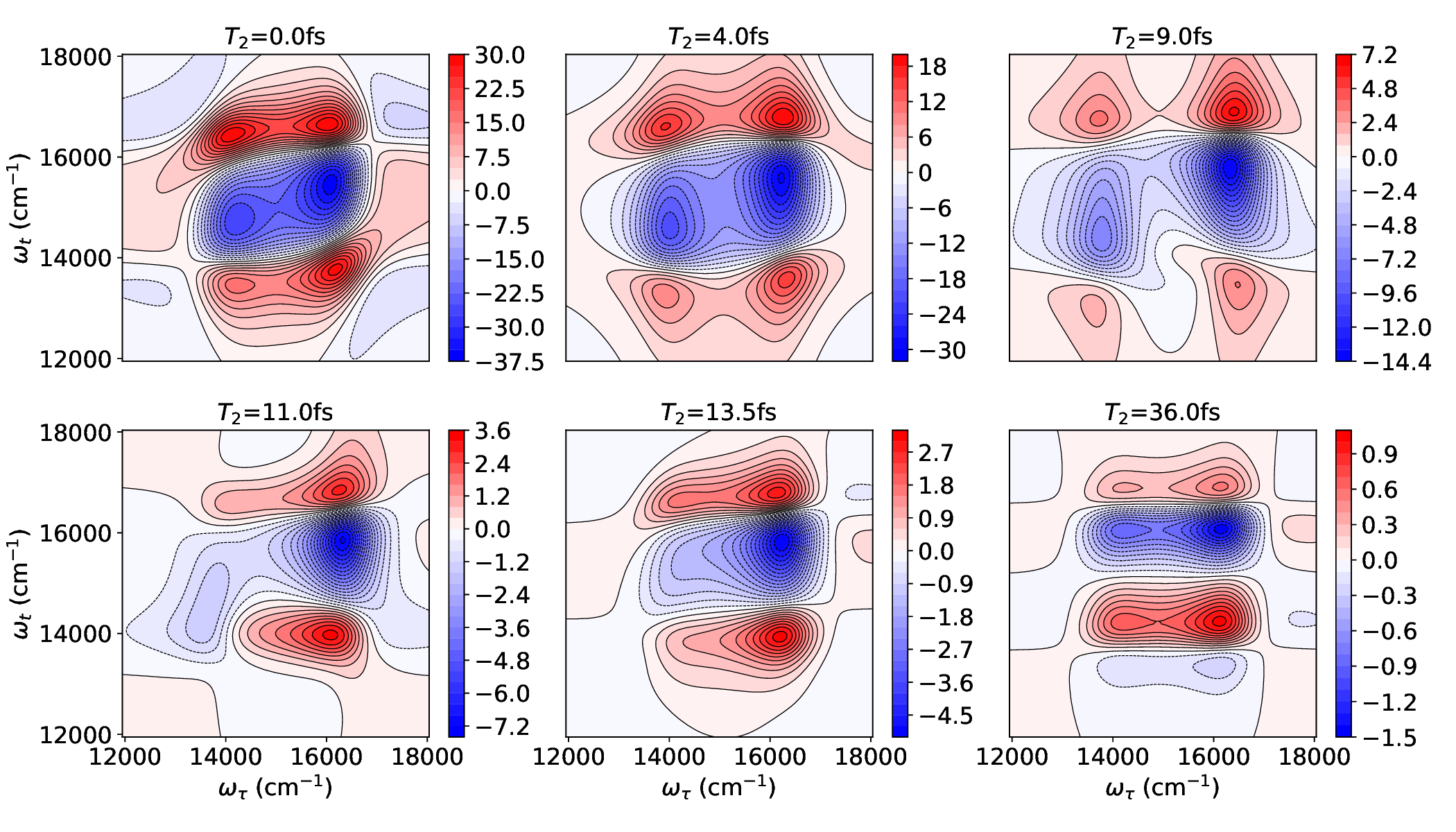}
    \vspace{1em}
    \caption{
    Simulated 2DES of a TLS coupled simultaneously to the plasmon
    mode and intramolecular vibrational DOFs for the strong
    coupling case $\alpha = 0.1$. All the other
    parameters are the same as those in Fig. \ref{fig-2d-0.1}.
    }
    \vspace{10em}
    \label{fig-2d-2bath-0.1}
\end{figure}

\begin{figure}[htbp]
    \centering
    \includegraphics[width=16cm]{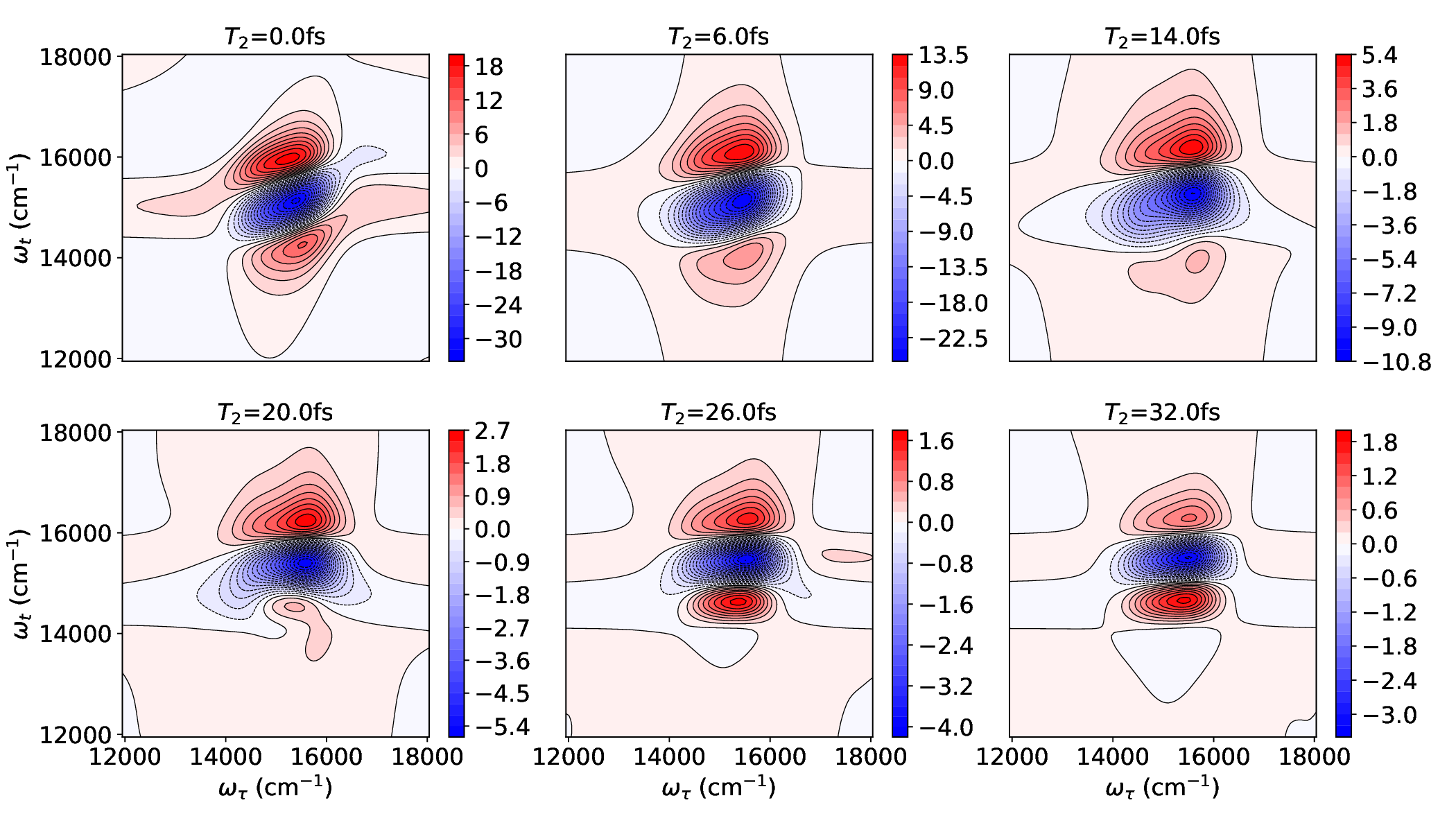}
    \vspace{1em}
    \caption{
    Same as Fig. \ref{fig-2d-2bath-0.1} for the intermediate 
    coupling case $\alpha = 0.05$.
    }
    \vspace{10em}
    \label{fig-2d-2bath-0.05}
\end{figure}

Finally, we investigate the effect of coupling the TLS to
intramolecular vibrational modes. The results for $\alpha =
0.1$ and $\alpha = 0.05$ are shown in Figs.
\ref{fig-2d-2bath-0.1} and \ref{fig-2d-2bath-0.05},
respectively. The parameters for the vibrational mode
coupling are the same as those used in Fig.
\ref{fig-abs-coupling}. The results show patterns similar to
those observed without coupling to the vibrational DOFs, 
except at longer waiting times. In the
strong coupling regime, the ``breathing mode" pattern remains
clearly visible, although the beating period is slightly
shorter. Additionally, the peak splitting at the longer
time, $T_2 = 36.0 \ \mathrm{fs}$, is less pronounced
compared to Fig. \ref{fig-2d-0.1}. For the intermediate
coupling case shown in Fig. \ref{fig-2d-2bath-0.05}, the
results are similar to those in Fig. \ref{fig-2d-0.05}, with
the exception of the peaks at longer waiting times, $T_2 =
26.0 \ \mathrm{fs}$ and $32.0 \ \mathrm{fs}$. Here, the
positive peak at smaller $\omega_t$ is more pronounced than
in Fig. \ref{fig-2d-0.05}.

\section{Discussion and conclusions}
\label{discussions}

In summary, we have developed a theoretical model to simulate energy
transfer dynamics, absorption spectra, and 2DES for a
quantum emitter coupled to a single plasmon mode. The
quantum emitter is described using a simplified TLS model.
The plasmon mode, due to its Bosonic nature and
short lifetime, is treated as a damped harmonic oscillator.
Acting as a bath to the TLS, it is further described by a Lorentzian
spectral density. The HEOM method is then employed to simulate
the dynamics and spectroscopic properties of the coupled
TLS-plasmon system over a wide range of parameters.

The absorption spectra of the plexciton system, both without
and with the effect of coupling the TLS to intramolecular
vibrational modes, are presented. The results indicate that,
increasing the TLS-plasmon coupling strength causes the
absorption spectra to transition from an asymmetric Fano line 
shape to a Rabi splitting pattern. 
Coupling the TLS to intramolecular vibrational DOFs reduces the
central dip in the absorption spectra, and results in a more
symmetric line shape. In the weak TLS-plasmon coupling
regime, this coupling may also suppress the asymmetric
Fano resonance feature, and lead to a single
absorption peak. Changing the dipole moment of the TLS does not
affect the peak splitting, but instead modifies the intensity
ratio of the lower and upper peaks.

We have also simulated the 2DES of plexciton systems. The
simulated 2DES is markedly different from those of coupled
dimers used to study excited-state energy transfer in
molecular systems, highlighting the unique features of the
nonlinear response in plexciton systems. In the strong
TLS-plasmon coupling regime, a ``breathing mode" pattern is
observed for the split peaks along the $\omega_\tau$
direction, as a signature of Rabi oscillations. In contrast,
the 2DES exhibits individual positive and negative peaks in
the intermediate and weak TLS-plasmon coupling regimes, with
the peak shapes transitioning from asymmetric to more
symmetric as the waiting time increases.

These findings on the 2DES of the plexciton system are
consistent with the excited-state population dynamics of the
TLS, and suggest that 2DES can be used to investigate energy
exchange between the TLS and the plasmon mode. The
theoretical model developed in this work provides a
foundation for future studies of more complex dynamic
behaviors in plexciton systems.
\cite{balci13,zengin13,leng18a,gross18,bitton19}


\acknowledgments
This work is supported by NSFC (Grant Nos. 21933011,
22203098, and 22433006).


\providecommand{\latin}[1]{#1}
\makeatletter
\providecommand{\doi}
  {\begingroup\let\do\@makeother\dospecials
  \catcode`\{=1 \catcode`\}=2 \doi@aux}
\providecommand{\doi@aux}[1]{\endgroup\texttt{#1}}
\makeatother
\providecommand*\mcitethebibliography{\thebibliography}
\csname @ifundefined\endcsname{endmcitethebibliography}
  {\let\endmcitethebibliography\endthebibliography}{}

\pagebreak

\end{document}